\documentclass[10pt]{iopart}
\input{iopams.sty}
\usepackage{graphicx}
\usepackage[dvips]{color}
\newcommand{\be}{\begin{equation}}

\newcommand{\ee}{\end{equation}}
\newcommand{\bea}{\begin{eqnarray}}
\newcommand{\eea}{\end{eqnarray}}

\newcommand{\ha}{\frac{1}{2}}
\newcommand{\rr}{{\bf r}}
\newcommand{\q}[2]{ {\bf q}_{#1}^{#2}}
\newcommand{\setq}[2]{\{{\bf q}_{#1}^{#2}\}}
\newcommand{\coleps}{\epsilon_{I\alpha\beta}}
\newcommand{\flaeps}{\epsilon_{Iij}}

\begin{document}
%\article[Crystallography of SQM]{SQM 2006}{The crystallography of strange quark matter}
\article[Crystallography of SQM]{}{The crystallography of strange quark matter
}
\date{\today}
\author{K. Rajagopal and R. Sharma\footnote{Speaker.}}
\address{Center for Theoretical Physics, Massachusetts Institute
of Technology, Cambridge, MA 02139}
\address{Nuclear Science Division, MS 70R319,
Lawrence Berkeley National Laboratory, Berkeley, CA 94720}
%\ead{\mailto{krishna@lns.mit.edu},\mailto{sharma@mit.edu}}

\begin{abstract}
Cold three-flavor quark matter
at large (but not asymptotically large) densities may exist as a crystalline
color superconductor. We explore this possibility
by calculating the gap parameter
$\Delta$ and free energy $\Omega(\Delta)$ for possible crystal structures
within a Ginzburg-Landau approximation, evaluating $\Omega(\Delta)$ to order
$\Delta^6$. We develop a qualitative understanding of what makes a
crystal structure stable, and find two structures 
with particularly large values of $\Delta$
and the condensation energy,
within a
factor of two of those for the CFL phase known to characterize
QCD at asymptotically large densities.  
The robustness of these phases
results in their being favored over wide ranges of density and though it
also implies that the Ginzburg-Landau approximation is not quantitatively
reliable,  previous work suggests that it can be trusted
for qualitative comparisons between crystal structures.
We close with a look ahead at the calculations that remain to
be done in order to make contact with observed
pulsar glitches and neutron star cooling.
\end{abstract}

We explore properties of quark matter at zero temperature and at densities that
may occur within neutron star cores. Absent interactions, each flavor of quark will fill momentum eigenstates up to a Fermi surface. However, this configuration is unstable to
the formation of Cooper pairs in the presence of any attractive interaction
between quarks. In QCD, the interaction between pairs of quarks that are antisymmetric
in color is attractive, and hence the ground state features a diquark condensate that is
dominantly antisymmetric in color. This is color superconductivity. We 
consider condensates antisymmetric in Dirac indices and, consequently, in
flavor also, implying that quarks in a pair have different flavors.

At asymptotic densities ($M_s/\mu\rightarrow 0$), where 
the three light quarks can be treated as massless, quark matter exists in
the CFL phase~\cite{Alford:1998mk} in which quarks of all three colors and all three
flavors form Cooper pairs with zero total momentum, yielding $\langle ds \rangle$,
$\langle us \rangle$ and $\langle ud \rangle$ condensates, and in which 
all fermionic excitations are gapped, with a gap parameter $\Delta_0\sim 10-100$~MeV.
However, at densities relevant for neutron star phenomenology, meaning
quark chemical potentials at most $\mu\sim 500$ MeV, the strange quark mass $M_s$
cannot be neglected. 
In neutral unpaired quark matter in weak equilibrium, $M_s$ introduces 
splitting between the Fermi surfaces for quarks of different flavor which
can be taken into account to lowest order in $M_s^2/\mu^2$
by treating the quarks as if they were massless but with chemical
potential splittings $\delta\mu_2\equiv (\mu_u-\mu_s)/2$ and 
$\delta\mu_3\equiv (\mu_d -\mu_u)/2$ given 
by $\delta\mu_2=\delta\mu_3\equiv \delta\mu = M_s^2/(8\mu)$.
Note that the splitting between unpaired Fermi surfaces increases
with decreasing density.
In the CFL phase, the Fermi momenta are {\it not} given by these optimal
values for unpaired quark matter; instead, the system pays a free energy price 
$\propto \delta\mu^2 \mu^2$
to equalize all Fermi momenta and gains a pairing energy benefit
$\propto \Delta_0^2\mu^2$.   As a function of decreasing density, there
comes a point (at which $\delta\mu \approx \Delta_0/4$~\cite{Alford:2003fq}) when
the system can lower
its energy by breaking pairs.  Below this density the 
CFL phase can certainly not be the ground
state of matter. Within a spatially homogeneous pairing ansatz, the phase
that results when CFL Cooper pairs start to break is the gapless
CFL (gCFL) phase of Ref.~\cite{Alford:2003fq}.  However, this
phase turns out to be unstable to the formation of counter-propagating
currents, presumably leading to inhomogeneity,
and therefore cannot be the ground state of matter at any density~\cite{Casalbuoni:2004tb}.  
Within the range of densities in which the gCFL phase has lower
free energy than the CFL and unpaired phases, the true ground state
of dense matter must have lower free energy still.  We have recently proposed 
a candidate phase (actually, two candidate phases) for the ground state
of matter over much of this density regime and at still lower densities~\cite{Rajagopal:2006ig}.
In these proceedings we describe this proposal. For further references,
in particular  to other suggested resolutions of the gCFL instability, 
see Ref.~\cite{Rajagopal:2006ig}.

Crystalline color superconductivity~\cite{Alford:2000ze,Bowers:2002xr} 
naturally permits pairing between quarks living at
split Fermi surfaces by allowing Cooper pairs with nonzero net momentum.
In three-flavor quark matter, this allows pairing to occur even with the
Fermi surfaces split in the free-energetically optimal way as in the absence
of pairing~\cite{Casalbuoni:2005zp,Mannarelli:2006fy,Rajagopal:2006ig}.  
This is the origin of the advantage that crystalline color superconducting
phases have over the CFL and gCFL phases at large values of the splitting $\delta\mu$.
For example, by
allowing $u$ quarks with momentum ${\bf p+q}_3$ to pair with $d$
quarks with momentum ${\bf -p+q}_3$, for any ${\bf p}$, 
we can pair $u$ and $d$ quarks
along rings on their respective Fermi surfaces~\cite{Alford:2000ze,Bowers:2002xr}. In coordinate
space, this corresponds to a condensate of the form 
$\langle ud \rangle\sim \Delta_3 \exp\bigl({2i{\bf q_3}\cdot{\bf r}}\bigr)$. 
The net free energy gained due to pairing is then a balance between increasing $|{\bf q}_3|$
yielding pairing on larger rings while exacting a greater kinetic energy cost. The optimum
choice turns out to be $|{\bf q}_3|=\eta \delta\mu_3$ with $\eta=1.1997$, corresponding
to pairing rings on the Fermi surfaces with opening angle $67.1^\circ$~\cite{Alford:2000ze}.
It is possible to cover larger areas of the Fermi surfaces by allowing Cooper pairs with 
the same $|{\bf q}_3|$ but various $\hat{\bf q}_3$, yielding 
$\langle ud \rangle \sim \Delta_3 \sum_{\q{3}{a}}\exp\bigl(2\, i\, \q{3}{a} \cdot {\bf r}\bigr)$
with the $\q{3}{a}$ chosen from some specified set $\setq{3}{}$.  This is a condensate
modulated in position space in some crystalline pattern, with the crystal structure defined
by $\setq{3}{}$.  In this two-flavor context, a Ginzburg-Landau
analysis reveals that 
the best $\setq{3}{}$ contains eight vectors pointing at the corners of a cube,
say in the $(\pm 1,\pm 1,\pm 1)$ directions in momentum space, yielding a face-centered
cubic structure in position space~\cite{Bowers:2002xr}.

We use the following ansatz for the three-flavor crystalline color superconducting 
condensate~\cite{Rajagopal:2006ig}:
\begin{equation}
\langle\psi_{i\alpha}C\gamma^5\psi_{j\beta}\rangle\propto
\sum_I\coleps\flaeps\Delta_I\sum_{{\bf q}_I^a}\exp({2i{\bf q}_I^a\cdot{\bf r}})\ .
\label{condensate}
\end{equation}
This is antisymmetric in color ($\alpha,\beta$), spin, 
and flavor ($i,j$) indices and is thus a generalization
of the CFL condensate to crystalline color superconductivity. 
We set $\Delta_1=0$, neglecting
$\langle ds \rangle$ pairing because the $d$ and $s$ Fermi
surfaces  are twice as far apart from each other as each is from
the intervening $u$ Fermi surface. Hence, $I$ can be taken to run over $2$ and $3$ only.
$\setq{2}{}$ and $\setq{3}{}$ define the crystal structures of the $\langle us \rangle$ and
$\langle ud \rangle$ condensates respectively.   We only consider crystal structures
in which all the vectors in $\setq{2}{}$ are equivalent to each other, and same for
$\setq{3}{}$, as this justifies our simplifying assumption that the $\langle us\rangle$
and $\langle ud \rangle$ condensates are each specified by a single gap parameter
($\Delta_2$ and $\Delta_3$ respectively), avoidin having to introduce one gap
parameter per ${\bf q}$.  We furthermore only consider crystal structures which
are exchange symmetric, meaning that $\setq{2}{}$ and $\setq{3}{}$ can be exchanged
by some combination of rigid rotations and reflections applied simultaneously to
all the vectors in both sets.  This simplification, together with 
$\delta\mu_2=\delta\mu_3$ (an approximation corrected only at order $M_s^4/\mu^3$),
guarantees that we find solutions with $\Delta_2=\Delta_3$.

We analyze and compare candidate crystal structures
by evaluating the free energy $\Omega(\Delta_2,\Delta_3)$ for each
crystal structure in
a Ginzburg-Landau expansion in powers of the $\Delta$'s. 
This approximation
is controlled
if $\Delta_2,\Delta_3 \ll \Delta_0,\delta\mu$, with $\Delta_0$ the gap parameter in the
CFL phase at $M_s^2/\mu=0$.  
The terms in the Ginzburg-Landau expansion must respect the global $U(1)$ symmetry 
for each flavor,
%. That is, $\Omega$ must be
%invariant under $\Delta_I \rightarrow e^{i\phi_I} \Delta_I$, 
meaning that each $\Delta_I$ can only
appear in the combination $|\Delta_I|^2$. (The $U(1)$ symmetries are spontaneously
broken by the condensate, but not explicitly broken.) Therefore,
$\Omega(\Delta_2,\Delta_3)$ is given to sextic order by
\begin{eqnarray}
\nonumber
\Omega(\Delta_2,\Delta_3)&=&\frac{2\mu^2}{\pi^2}\Biggl[P_2
\alpha_2 |\Delta_2|^2 + P_3 \alpha_3 |\Delta_3|^2
+\ha\Bigl( \beta_2|\Delta_2|^4 + \beta_3|\Delta_3|^4
+ \beta_{32} |\Delta_2|^2|\Delta_3|^2\Bigr)\\
&+&\frac{1}{3}\Bigl( \gamma_2|\Delta_2|^6+\gamma_3|\Delta_3|^6
+\gamma_{322}|\Delta_3|^2|\Delta_2|^4+\gamma_{233}|\Delta_3|^4|\Delta_2|^2\Bigr)\Biggr]
\label{GLexpansion}\;,
\end{eqnarray}
where we have chosen notation consistent with that  used in
the two flavor study of Ref.~\cite{Bowers:2002xr}, which arises as a special
case of (\ref{GLexpansion}) if we take $\Delta_2$ or $\Delta_3$ to be zero. 
$P_I$ is the number of vectors in the set $\setq{I}{}$. 
The form of the
Ginzburg-Landau expansion (\ref{GLexpansion}) is model-independent, whereas
the expressions for the coefficients $\alpha_I$, $\beta_I$, $\beta_{IJ}$, $\gamma_I$,
and $\gamma_{IJJ}$ for a specific crystal structure are model-dependent.
For exchange symmetric
crystal structures, $\alpha_2=\alpha_3\equiv\alpha$, $\beta_2=\beta_3\equiv\beta$, 
$\gamma_2=\gamma_3\equiv\gamma$ and $\gamma_{233}=\gamma_{322}$.

Because setting one of the $\Delta_I$ to zero reduces the problem to one with two-flavor
pairing only, we can obtain $\alpha$, $\beta$ and $\gamma$
via applying the two-flavor analysis described in Ref.~\cite{Bowers:2002xr} to
either $\setq{2}{}$ or $\setq{3}{}$ separately.  
Using $\alpha$ as an example, we learn 
that
\begin{equation}
\alpha_I =
\alpha(q_I,\delta\mu_I)= -1
 +\frac{\delta\mu_I}{2 q_I}
 \log\left(\frac{q_I+\delta\mu_I}{q_I-\delta\mu_I}\right)
 -\ha\log\left(\frac{\Delta_{\rm 2SC}^2}{4(q_I^2-\delta\mu_I^2)}\right)\ ,
\label{AlphaEqn}
\end{equation}
Here, $q_I\equiv |{\bf q}_I|$ and 
$\Delta_{\rm 2SC}$ is the gap parameter
for the 2SC (2-flavor, 2-color) BCS pairing 
obtained with $\delta\mu_I=0$ and $\Delta_I$ nonzero with the
other two gap parameters set to zero.    Assuming that $\Delta_0\ll \mu$, it
is given by
$\Delta_{\rm 2SC}= 2^{\frac{1}{3}}\Delta_0$~\cite{Schafer:1999fe}.
In the Ginzburg-Landau approximation, in which the $\Delta_I$ are assumed
small, we must  first minimize the quadratic contribution to the
free energy, and only then  investigate the 
quartic and sextic contributions.  Minimizing $\alpha_I$ fixes the length
of all the vectors in the set $\{{\bf q}_I\}$, yielding
$q_I = \eta\, \delta\mu_I$ with $\eta=1.1997$ the solution to
$\frac{1}{2\eta}\log\left[(\eta+1)/(\eta-1)\right]=1$~\cite{Alford:2000ze}. 
Upon setting $q_I=\eta\,\delta\mu_I$, (\ref{AlphaEqn}) becomes
\begin{equation}
\alpha_I(\delta\mu_I) 
=-\frac{1}{2}
\log\left(\frac{\Delta_{\rm 2SC}^2}{4 \delta\mu_I^2 (\eta^2-1)}\right)\ .
\label{AlphaEqn2}
\end{equation}
Furthermore, the only dimensionful
quantities on which the quartic and sextic coefficients can depend are 
then the $\delta\mu_I$~\cite{Bowers:2002xr,Rajagopal:2006ig},
meaning that for exchange symmetric crystal structures and with $\delta\mu_2=\delta\mu_3=\delta\mu$
we have 
$\beta=\bar\beta/\delta\mu^2$, $\beta_{32}=\bar\beta_{32}/\delta\mu^2$, 
$\gamma=\bar\gamma/\delta\mu^4$ and $\gamma_{322}=\bar\gamma_{322}/\delta\mu^4$
where the barred quantities are dimensionless numbers which depend only on
$\{\hat{\bf q}_2\}$ and $\{\hat{\bf q}_3\}$ that must  be evaluated for each crystal structure.
Doing so
%(described for $\bar\beta$ and $\bar\gamma$ in Ref.~\cite{Bowers:2002xr}
%and for $\bar\beta_{32}$ and $\bar\gamma_{322}$ in Ref.~\cite{Rajagopal:2006ig}) 
requires
evaluating one loop Feynman diagrams with 4 or 6 insertions of $\Delta_I$'s. Each insertion
of $\Delta_I$ ($\Delta_I^*$) adds (subtracts) momentum $2\q{I}{a}$ for some $a$, meaning that
the calculation consists of a bookkeeping task (determining which combinations of
4 or 6 $\q{I}{a}$'s are allowed) 
that grows rapidly in complexity with the compexity of the crystal structure
and a loop integration that is nontrivial because the momentum in the propagator
changes after each insertion. See Ref.~\cite{Rajagopal:2006ig}, where this calculation
is carried out explicitly for 11 crystal structures in a mean-field NJL model upon
making the weak coupling ($\Delta_0$ and $\delta\mu$ both much less than $\mu$)
approximation. Note that in this approximation
neither the NJL cutoff nor the NJL coupling constant appear in any quartic
or higher Ginzburg-Landau coefficient, and they appear in $\alpha$ only 
within  $\Delta_0$.
Hence, the details of the model do not matter as long as one thinks of 
$\Delta_0$ as a parameter, kept $\ll \mu$.

It is easy to show that  for exchange symmetric crystal structures any
extrema of $\Omega(\Delta_2,\Delta_3)$ in $(\Delta_2,\Delta_3)$-space
must either have $\Delta_2=\Delta_3=\Delta$,
or have one of $\Delta_2$ and $\Delta_3$ vanishing~\cite{Rajagopal:2006ig}.  
It is also possible to show that 
the three-flavor crystalline phases with 
$\Delta_2=\Delta_3=\Delta$ are electrically neutral whereas two-flavor solutions
in which only one of the $\Delta$'s is nonzero are not~\cite{Rajagopal:2006ig}.  
We therefore analyze only solutions with $\Delta_2=\Delta_3=\Delta$.
In marked contrast to the two-flavor results of Ref.~\cite{Bowers:2002xr}
which (ignoring the requirement of neutrality)
show that many two-flavor crystal structures have negative $\gamma$ and hence
sextic order free energies that are unbounded from below, 
we find that $\Omega(\Delta,\Delta)$ is positive for large $\Delta$
for all the crystal structures that we investigate~\cite{Rajagopal:2006ig}.
This allows us to  minimize $\Omega(\Delta,\Delta)$ with respect to $\Delta$,
thus evaluating $\Delta$ and $\Omega$.

We begin with the simplest three-flavor  ``crystal'' structure in which
$\setq{2}{}$ and $\setq{3}{}$ each contain only a single vector, making
the $\langle us\rangle$ and $\langle ud\rangle$ condensates 
each a single plane wave.  
This simple condensate 
yields a qualitative lesson which proves helpful in winnowing the space
of multiple plane wave crystal structures~\cite{Rajagopal:2006ig}.
For this simple ``crystal'' structure, all the coeffficients in the Ginzburg-Landau
free energy can be evaluated analytically~\cite{Rajagopal:2006ig}. The 
terms that occur in the three-flavor case but not in the two-flavor case,
namely $\bar\beta_{32}$ and $\bar\gamma_{322}$, describe the
interaction between the two condensates, and depend on the angle
$\phi$ between $\q{2}{}$ and $\q{3}{}$. 
For any angle $\phi$, both $\bar{\beta}_{32}$ and
$\bar{\gamma}_{322}$ are
positive.
And, both increase monotonically with $\phi$ and
diverge as $\phi\rightarrow\pi$~\cite{Rajagopal:2006ig}.
%This tells us that within this two plane
%wave ansatz, the most favorable orientation is $\phi=0$, namely
%$\q{2}{}\parallel\q{3}{}$.  
This divergence 
tells us that choosing $\q{2}{}$ and
$\q{3}{}$ precisely antiparallel exacts an infinite free energy price
in the combined Ginzburg-Landau and weak-coupling
limit in which $\Delta\ll \delta\mu,\Delta_0\ll \mu$,
meaning that in this limit if we chose $\phi=\pi$ we find $\Delta=0$.
Away from the Ginzburg-Landau limit, when the pairing rings
on the Fermi surfaces widen into bands, choosing $\phi=\pi$ exacts a finite
price meaning that $\Delta$  is nonzero but smaller than that for any other
choice of $\phi$.  
The high cost of choosing $\q{2}{}$ and
$\q{3}{}$ precisely antiparallel can be understood 
qualitatively as arising from the fact that in this case the
ring of states on the $u$-quark Fermi surface that ``want to'' pair
with $d$-quarks coincides precisely with the ring that ``wants to''
pair with $s$-quarks~\cite{Mannarelli:2006fy}.  
This simple two plane wave ansatz has been
analyzed upon making
the weak-coupling approximation but without
making the Ginzburg-Landau approximation~\cite{Mannarelli:2006fy}.
All the qualitative lessons learned from the Ginzburg-Landau approximation
remain valid
and we learn further that 
the Ginzburg-Landau 
approximation always underestimates $\Delta$~\cite{Mannarelli:2006fy}.

The analysis of the simple two plane wave ``crystal'' structure, together with
the observation that in more complicated crystal structures with more than
one vector in $\setq{2}{}$ and $\setq{3}{}$ the Ginzburg-Landau coefficient
$\beta_{32}$ ($\gamma_{322}$) is given in whole (in part) by a sum of many 
two plane wave contributions,
yields one of two
rules for constructing favorable crystal structures for three-flavor
crystalline color superconductivity~\cite{Rajagopal:2006ig}:
$\setq{2}{}$ and $\setq{3}{}$ should be rotated with respect to each other
in a way that best keeps vectors in one set away from the antipodes
of vectors in the other set.  
The second rule is that 
the sets $\setq{2}{}$ and $\setq{3}{}$ should each
be chosen to yield crystal structures which, seen
as separate two-flavor crystalline phases, are as favorable as possible.
%Recall that
%the favored 
%$\setq{2}{}$ or $\setq{3}{}$ in isolation consists of eight
%vectors pointing at the corners of a cube~\cite{Bowers:2002xr}.  
The 11 crystal structures analyzed in Ref.~\cite{Rajagopal:2006ig}
allow one to
make several pairwise
comparisons that test these two rules.
There are instances of two structures which differ only in the
relative orientation of $\setq{2}{}$ and $\setq{3}{}$ and in these cases the
structure in which vectors in $\setq{2}{}$ get closer to the antipodes
of vectors in $\setq{3}{}$ are disfavored. And, there are instances
where the smallest angle between a vector in $\setq{2}{}$ and the antipodes
of a vector in $\setq{3}{}$ are the same for two different crystal structures,  and
in these cases the one with the more favorable two-flavor structure is more favorable.
These considerations, together with explicit calculations, indicate that
two structures, which we denote ``2Cube45z'' and ``CubeX'', are particularly favorable.

In the 2Cube45z crystal, $\setq{2}{}$ and $\setq{3}{}$ each contain eight
vectors pointing at the corners of a cube. If we orient $\setq{2}{}$ so that its
vectors point in the $(\pm 1,\pm 1,\pm 1)$ directions in momentum space, then $\setq{3}{}$ is rotated
relative to $\setq{2}{}$ by $45^\circ$ about the $z$-axis.   In this crystal structure,
the $\langle ud \rangle$ and $\langle us \rangle$ condensates are each
given by the most favored 
two-flavor crystal structure~\cite{Bowers:2002xr}.  The relative rotation
maximizes the separation between any vector in
$\setq{2}{}$ and the nearest antipodes of a vector in $\setq{3}{}$.

\begin{figure}[t]
  \begin{center}
    \begin{minipage}[c]{0.5\linewidth}
     \caption{{ The CubeX crystal structure of Eq.~(\ref{CubeXStructure}). 
The figure extends from 0 to $a/2$ in the $x$, $y$ and $z$ directions.
Both $\Delta_2(\rr)$ and $\Delta_3(\rr)$
vanish at the horizontal plane.  $\Delta_2(\rr)$ vanishes on the darker vertical planes,
and $\Delta_3(\rr)$ vanishes on the lighter vertical planes.   
On the upper
(lower) dark cylinders and the lower (upper) two small corners of dark cylinders,
$\Delta_2(\rr)=  +3.3 \Delta$ ($\Delta_2(\rr)=  -3.3 \Delta$).
On the upper
(lower) lighter cylinders and the lower (upper) two small corners of lighter cylinders,
$\Delta_3(\rr)=  -3.3 \Delta$ ($\Delta_3(\rr)=  +3.3 \Delta$).  
The largest value of $|\Delta_I(\rr)|$ is $4\Delta$, occurring along lines 
at the centers of the cylinders.
The lattice spacing is $a$ when one takes into
account the signs of the condensates; 
if one looks only at $|\Delta_I(\rr)|$, 
the lattice spacing is $a/2$. 
%%In (\ref{CubeXStructure}) and
%%hence in this figure, we have made a particular choice for the relative position of
%%$\Delta_3(\rr)$ versus $\Delta_2(\rr)$.  We show in Appendix B that one can be
%%translated relative to the other with no cost in free energy. 
}\label{CubeXContours}}
\vspace{-0.2in}
    \end{minipage}\hfill
    \begin{minipage}[c]{0.45\linewidth}
\vspace{-0.2in}
 \includegraphics[width=6.0cm,angle=0]{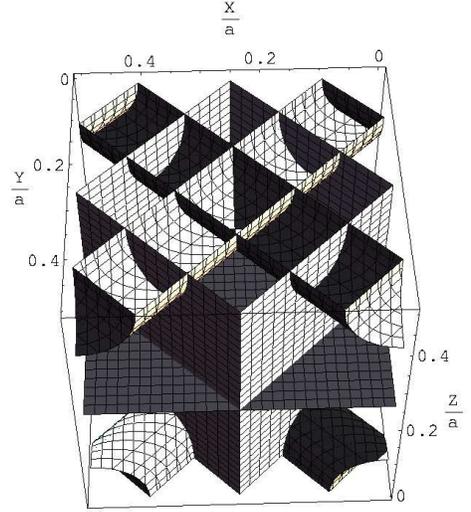} 
\vspace{-0.2in}
    \end{minipage}
  \end{center}
\end{figure}

We arrive at the CubeX structure by reducing the number
of vectors in $\setq{2}{}$ and $\setq{3}{}$.  This worsens the two-flavor
free energy of each condensate separately, but allows vectors in 
$\setq{2}{}$ to be kept farther away from the antipodes of vectors
in $\setq{3}{}$.  We have not analyzed all structures
obtainable in this way, but we have found one and only one which
has a condensation energy comparable to that of the 2Cube45z structure.
In the CubeX structure,
$\setq{2}{}$ and $\setq{3}{}$ each contain four vectors forming a rectangle. The eight 
vectors together point toward the corners of a cube. The 2 rectangles intersect to look like an ``X'' 
if viewed end-on.  
The color, flavor and position space dependence of 
the CubeX condensate is given by
\begin{eqnarray}
\epsilon_{2\alpha\beta}&&\epsilon_{2ij} \, 
2\Delta \Biggl[ 
\cos \frac{2\pi}{a} \left( x+y+z\right) + \cos \frac{2\pi}{a}\left(-x-y+z\right) \Biggr]\nonumber\\
&&+ \epsilon_{3\alpha\beta}\epsilon_{3ij} \, 2\Delta \Biggl[
\cos \frac{2\pi}{a} \left( -x+y+z\right) + \cos \frac{2\pi}{a}\left(x-y+z\right) \Biggr]\ ,
\label{CubeXStructure}
\end{eqnarray}
where
$a = \sqrt{3}\pi/q = 4.536/\delta\mu = \mu /(1.764 M_s^2)$
is the lattice spacing.  For
example, with $M_s^2/\mu=100, 150, 200$~MeV the lattice
spacing is  $a=72, 48, 36$~fm. 
We depict this condensate in Fig.~\ref{CubeXContours}.

In Figs.~2 and 3,
we plot $\Delta$
and $\Omega$ versus $M_s^2/\mu$ 
for the most favorable crystal structures that we have found, namely the
CubeX and 2Cube45z structures described above.
We have taken the CFL gap parameter $\Delta_0=25$~MeV in these figures,
but they can easily be rescaled to any value of $\Delta_0\ll \mu$~\cite{Rajagopal:2006ig}.
Fig.~2 shows that the gap parameters 
are large enough that the Ginzburg-Landau approximation is
at the edge of its domain of reliability.
However, results obtained for simpler crystal structures
suggest that the Ginzburg-Landau calculation underestimates $\Delta$ and the
condensation energy and that, even when
it breaks down, it is a good qualitative guide
to the favorable structure~\cite{Mannarelli:2006fy}.  We therefore trust the
result, evident in Fig.~3, that these crystalline phases are both
impressively robust, with one or other of them favored over a wide swath of
$M_s^2/\mu$ and hence density. We do not trust the Ginzburg-Landau calculation
to discriminate between these two 
structures, particularly given that although we have a qualitative understanding
of why these two are favorable we have no qualitative argument for why one should
be favored over the other. We are confident that 2Cube45z
is the most favorable structure obtained by rotating one 
cube relative to another. We are not as confident that CubeX is the best possible
structure with fewer than 8+8 vectors. Regardless, the 2Cube45z and CubeX crystalline
phases together make the case
that three-flavor crystalline color superconducting phases are the ground
state of cold quark matter over a wide range of  densities.  If 
even better crystal structures can be found, this will only further strengthen this case.

Fig.~3 shows that over most of the range of $M_s^2/\mu$ where it was once considered
a possibility, the gCFL phase can be replaced by a {\it much} more favorable three-flavor
crystalline color superconducting phase.  It also
shows that it is hard to find a crystalline
phase with lower free energy than the gCFL phase at the lowest values
of $M_s^2/\mu$ (highest densities) in the ``gCFL window''.  
This narrow window where the gCFL curve
remains the lowest in Fig.~3 is thus the most likely place
in the QCD phase diagram to find the gCFL phase augmented
by current-carrying meson condensates described in 
Refs.~\cite{Kryjevski:2005qq}.
Except within this window, crystalline color superconducting
phases with the CubeX or the 2Cube45z crystal structures
provide an attractive resolution to the instability of the gCFL phase.

\begin{figure}[t]
\centering
\begin{tabular}{cc}
\begin{minipage}{3.in}
\includegraphics[width=3in,angle=0]{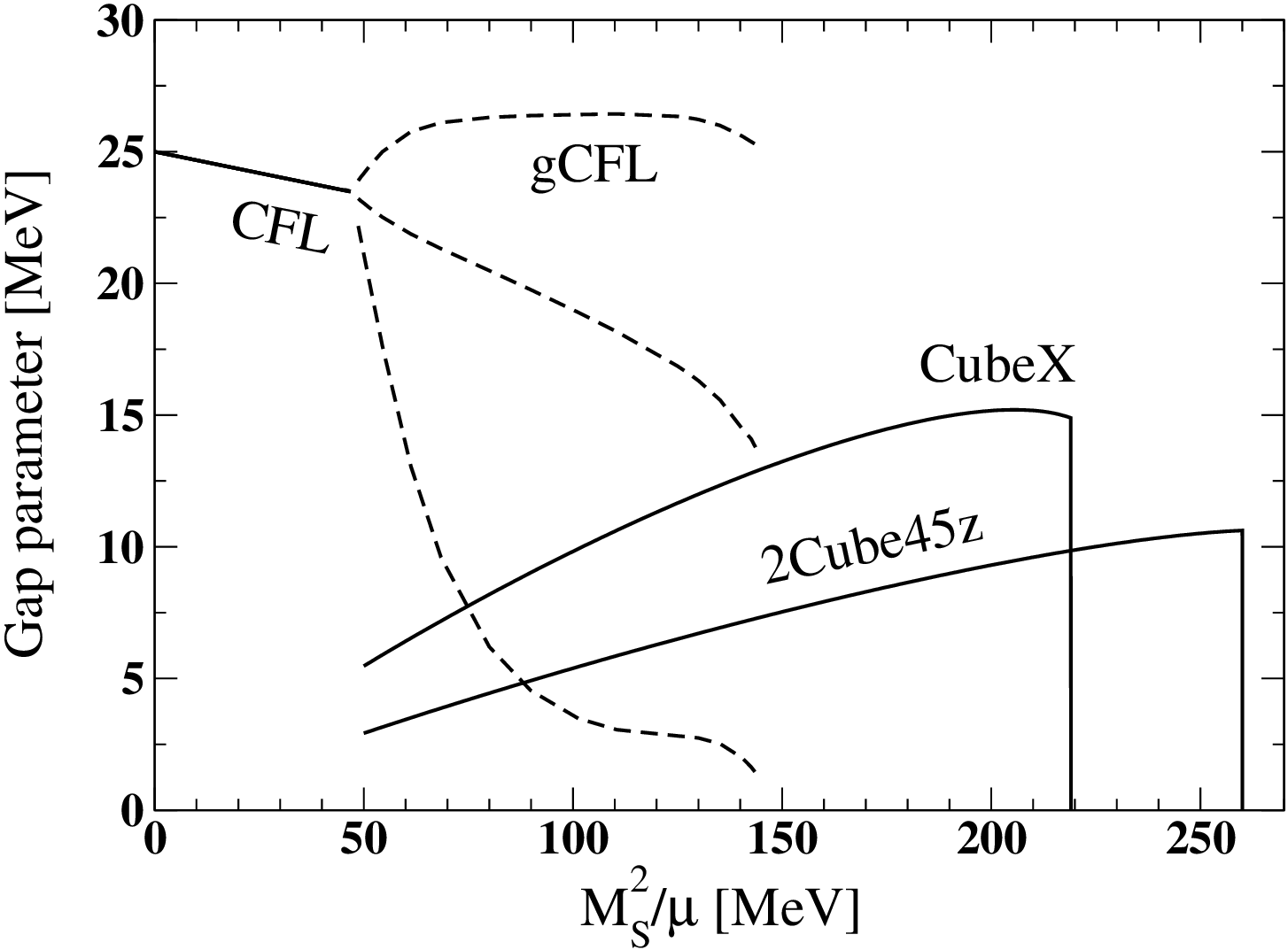}
\label{deltavsx}
\vspace{-0.2in}
\caption{Gap parameter $\Delta$ versus $M_s^2/\mu$ for the CFL gap parameter, the three gap
parameters $\Delta_1<\Delta_2<\Delta_3$ describing $\langle ds\rangle$,
$\langle us\rangle$ and $\langle ud\rangle$ pairing in the gCFL phase, and the
gap parameter in the
crystalline color superconducting phases with CubeX and 2Cube45z  crystal
structures. 
%Increasing $M_s^2/\mu$ corresponds to decreasing density.
}
\end{minipage}
&
\begin{minipage}{3.in}
\vspace{-0.05in}
\includegraphics[width=3in,angle=0]{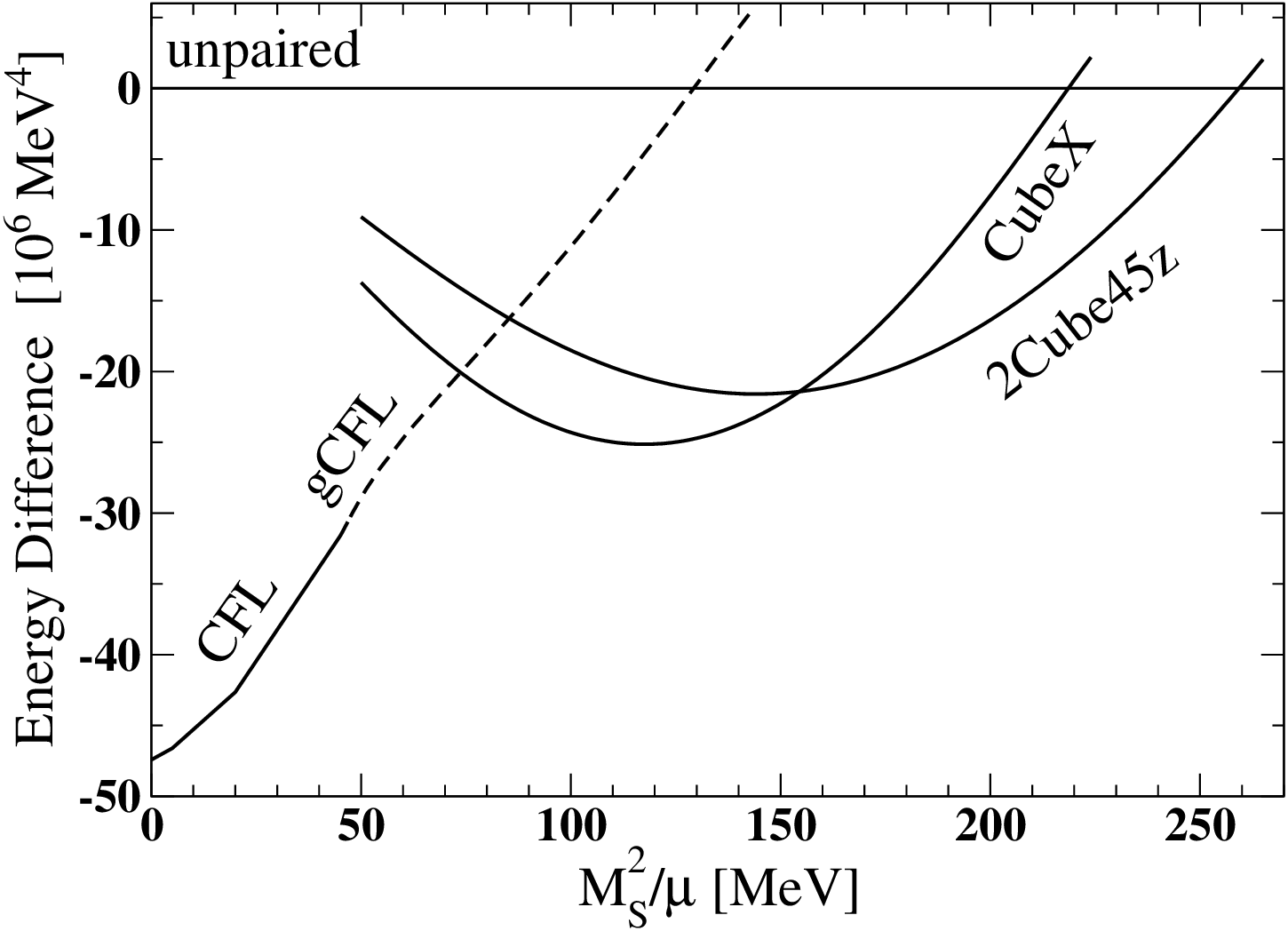}
\label{omegavsx}
\vspace{-0.2in}
\caption{Free energy $\Omega$ relative to that for neutral unpaired
quark matter versus $M_s^2/\mu$, for the same phases as in Fig.~2. 
%In both Figs~2 and 3, the interaction between quarks
%has been chosen such that the CFL gap parameter is $\Delta_0=25$~MeV at 
%$M_s^2/\mu=0$.  
The gCFL phase free energy is a dashed line
as a reminder that 
this phase is unstable: the free energy of the ground state of matter must lie below this
dashed line.}
\end{minipage} 
\end{tabular}
\end{figure}

The three-flavor crystallline color superconducting phases with 
CubeX and 2Cube45z crystal structures are
the lowest free energy phases that we know of, and hence
candidates for the ground state of QCD, over a wide range
of densities.  One or other is favored over the CFL, gCFL and
unpaired phases for $2.9 \Delta_0 < {M_s^2}/{\mu} < 10.4 \Delta_0$,
as shown in Fig.~3.
For $\Delta_0=25$~MeV and $M_s=250$~MeV,
this translates to $240 {\rm MeV} < \mu < 847 {\rm MeV}$.
With these choices of parameters, then, the 
lower part of this range of $\mu$ (higher part of the range of $M_s^2/\mu$) is certainly
superseded by nuclear matter.  And, the high end of this range extends far
beyond the $\mu\sim 500$~MeV characteristic of the quark matter at the densities
expected 
at the center of compact stars. If compact stars
have quark matter cores, then, it is reasonable to 
include 
the possibility that 
the {\it entire}
quark matter core could be in a crystalline color superconducting phase
on the menu of options
that must ultimately be winnowed by confrontation with astrophysical observations.
If $\Delta_0$ is larger, say $\sim 100$~MeV, the entire quark matter
core could be in the CFL phase.  

Now that we have two candidates
for the crystal structure of the  three-flavor crystalline color superconducting
phase of cold quark matter, favorable over a very wide range of intermediate
densities, and a qualitative
guide to the scale of $\Delta$ and $\Omega(\Delta)$, we can look ahead
toward the calculation of astrophysically relevant observables. 
The heat capacity of the CubeX and 2Cube45z phases should be only quantitatively
suppressed relative to that of unpaired quark matter, but their neutrino emissivity
may be more significantly suppressed~\cite{Rajagopal:2006ig}. Both
can now be calculated, yielding an estimate of
the effects of a crystalline quark matter 
core on the rate at which a neutron star  cools by neutrino emission. 
A crystalline color superconducting 
core, being both crystalline and superfluid, could be a region within which
rotational vortices are pinned and hence 
(some) pulsar glitches originate~\cite{Alford:2000ze}.  Or, the presence
of crystalline quark matter within neutron stars could be ruled out
if it predicts glitch phenomenology in qualitative disagreement with
that observed.  
The two microphysical properties of crystalline
quark matter that must be estimated before glitch phenomenology can
be addressed are the vortex pinning force and the shear modulus.  
Glitches occur if rotational
vortices are pinned and immobile while the spinning pulsar's angular velocity 
slows over years,  with the glitch triggered by the catastrophic unpinning
of long-immobile vortices.  Immobilizing vortices requires sufficient
pinning force and shear modulus.  
Estimating the pinning force
will require
analyzing how the CubeX and 2Cube45z phases respond when rotated. 
The shear modulus is
related to the coefficients in the effective theory that
describes the phonon modes of the crystal~\cite{Casalbuoni:2002my}.  
Analyzing the phonons in the three-flavor crystalline color superconducting phases
with the 2Cube45z and CubeX crystal structures is thus a priority.
These are the prerequisites to determining whether observations of
pulsar glitches can be used to rule out (or in) the presence of quark matter
in a crystalline color superconducting phase within neutron stars.

We thank the LBNL Nuclear Theory Group for hospitality. Research 
supported in part by the U.S.~DOE under contract
\#DE-AC02-05CH11231 and cooperative research agreement
\#DF-FC02-94ER40818.

\end{document}